\documentclass[12pt]{article}
\usepackage{epsfig, amssymb}
\usepackage{graphicx,epsfig}
\begin{document}
\begin{titlepage}
\thispagestyle{empty}
\begin{flushright}
\end{flushright}

\bigskip

\begin{center}
  \noindent{\Large \textbf
    {Stochastic quantization and holographic Wilsonian renormalization group }}\\

\vspace{2cm} \noindent{ Jae-Hyuk
  Oh${}^{a}$\footnote{e-mail:jack.jaehyuk.oh@gmail.com} and Dileep
  P. Jatkar${}^{a}$\footnote{e-mail:dileep@hri.res.in}}

\vspace{1cm}
{\it
Harish-Chandra Research Institute, \\
  Chhatnag Road, Jhunsi, Allahabad-211019, India${}^{a}$\\
}
\end{center}

\vspace{0.3cm}
\begin{abstract}
We study relation between stochastic quantization and holographic
  Wilsonian renormalization group flow. Considering stochastic
  quantization of the boundary on-shell actions with the Dirichlet
  boundary condition for certain $AdS$ bulk gravity theories, we find
  that the radial flows of double trace deformations in the boundary
  effective actions are completely captured by stochastic time
  evolution with identification of the $AdS$ radial coordinate `$r$' with
  the stochastic time `$t$' as $r=t$.  More precisely, we investigate
  Langevin dynamics and find an exact relation between radial
  flow of the double trace couplings and 2-point correlation functions
  in stochastic quantization.
  We also show that the radial evolution of double trace deformations
  in the boundary effective action and the stochastic time evolution
  of the Fokker-Planck action are the same.
  We demonstrate this relation with a couple of examples: (minimally
  coupled)massless scalar fields in $AdS_2$ and $U(1)$ vector fields
  in $AdS_4$.
\end{abstract}
\end{titlepage}

\newpage

\tableofcontents

\section{Introduction} {\it AdS/CFT} correspondence has shed a lot of
light on strongly coupled field theories.  Investigation of the
holographic renormalization group (RG) flows, for example, has become
a useful tool to understand the Wilsonian RG flow of strongly coupled
dual field theories.  In fact, with the recent improved understanding
of the holographic RG\cite{Akhmedov:1998vf,de Boer:1999xf}, it has
become clear that these two approaches to RG flow of the boundary
theory are consistent with each other\cite{Polchinski1,Hong1}.  In the
dual theory defined in $AdS$ space, $AdS$ radial coordinate $r$ is
identified with Wilsonian RG-direction, in other words, the radial
direction is related to the energy scale of the $CFT$. $AdS$
boundary($r=0$) is treated as the $UV$-region whereas Poincar\' e
horizon($r=\infty$) is treated as the $IR$-region.  For finite
nonvanishing values of $r$, one can define a $CFT$ at the intermediate
energy scale.

A method for computing holographic Wilsonian RG flows of certain
deformations of the theory defined on the $UV$ boundary was developed
in \cite{Polchinski1,Hong1}. (For earlier work on relevance of
multi-trace operators to holographic RG, see \cite{Akhmedov:2002gq}.)
The flow equation has a form of the Hamilton-Jacobi equation in the
limit when bulk action is restricted to the terms up to two
derivatives.  The most important feature of this computation is that
even though one has a free theory in the dual gravity, the flow
equations necessarily contain double trace deformations as long as
non-zero momenta along boundary directions are turned on. For zero
momentum case, the flow becomes rather trivial without these double
trace couplings.  The double trace deformation coupling has evoked a
lot of interest recently.  For example, for double trace couplings of
transverse(longitudinal) boundary $U(1)$ gauge fields appearing in the
boundary effective actions with bulk $U(1)$ gauge fields in $AdS_4$,
the equations of these couplings correspond to the flow equations of
transverse(longitudinal) conductivities in the dual fluid system
defined on $AdS$ boundary \footnote{There are many computations of
  transport coefficients using holographic Wilsonian RG(equivalently
  the sliding membrane paradigm), such as shear viscosity
  \cite{Oh:2012zu,Mamo:2012sy} and conductivities\cite{Iqbal:2008by,Sin:2011yh,Rebhan:2011vd,Ge:2011fb}.}.

The double trace couplings show several fixed points in $UV$ boundary,
which depend on the boundary conditions on it. In \cite{Hong1}, they
provide examples of the flow equations for bulk scalar fields with its
mass $m$, which is in the range that $-\frac{d^2}{4}\leq m^2\leq
-\frac{d^2}{4}+1$, where $d$ is spacetime dimension of the boundary
and for $U(1)$ gauge fields in $AdS_4$. The most important property of
both bulk theories is that they allow alternative
quantization\cite{Daniel1,Witten11,Klebanov:1999tb,Witten:2001ua,Ioannis1,Sebastian1}.  In
this case, one can impose both Neumann boundary condition as well as
Dirichlet boundary condition on the conformal boundary.  These
boundary conditions correspond to alternative and standard
quantization respectively and they lead to different $UV$ fixed
points. Moreover, there are many classes of flows which do not start
from fixed points in the $UV$ region.

However, in the $IR$ region, near Poincar\' e horizon, it turns out that
most of the flows converge to a single fixed point for these
examples\footnote{This is no longer true when the bulk geometry is
  that of an extremal black brane.  In that case, there is emergent
  1-dimensional $CFT$ near the black brane horizon and couplings of bulk
  fields admitting alternative quantization even in $AdS_2$, may give
  rise to other nontrivial fixed points.}(where background geometry of
the bulk is Poincar\' e patch of the pure $AdS$ space). Moreover, it
turns out that the boundary effective action in $IR$ region has the
same form as the classical effective action $\Gamma$.  $\Gamma$ is
derived from the on-shell action ${I_{os}}$ by Legendre transform,
where the on-shell action is obtained from bulk action by imposing
Dirichlet boundary condition at the $UV$ boundary.

There have been some attempts in the past trying to relate $AdS/CFT$
and stochastic
quantization\cite{Lifschytz:2000bj,Polyakov:2000xc,Petkou1,Minic:2010pw}.
Stochastic quantization\cite{Huffel1,Dijkgraaf1} is a quantization method for Euclidean field
theories where one starts with a $d$-dimensional Euclidean action,
$S_c(\phi)$(which is also called the classical action).  The coupling
of the field $\phi$ to the surrounding is mimicked by Gaussian white
noise $\eta$, which is the source of randomness or stochasticity in
the system.
Stochastic system evolves along the stochastic time $t$, which is
different from the Euclidean time, $\tau$.  At very late time
$t\rightarrow \infty$, the system settle down to an equilibrium state,
and partition function of it provides correlation functions of quantum
field theory with action $S_c$. Even if the system starts with
$d$-dimensional Euclidean action, the resulting theory is
$d+1$-dimensional since even in equilibrium the system is evolving
along the stochastic time `$t$'.  In fact, $AdS/CFT$ correspondence
has similar structure. Conformal field theories on the $d$-dimensional
$AdS$ boundary are related to $d+1$-dimensional bulk string theories
and the radial coordinate `$r$' in $AdS$ space has similar role to
play as the stochastic time `$t$'.

In particular, there is a rather concrete conjecture for the
relation\cite{Petkou1}, which basically depends on an identification
of a partition function derived from the holographic method
with the stochastic partition function. Holographic partition
function
is given by
\begin{equation}
Z_{hol}=\int_{\phi(r=0)=\phi_0} [D \phi]e^{-S_{bulk}(\phi)}=e^{-W(\phi_0)},
\end{equation}
where the boundary is $AdS$ boundary which is located at $r=0$, $\phi$
denotes the bulk field and $\phi_0$ is its boundary value. In the
above expression, we have imposed Dirichlet boundary condition at
$r=0$ and $W(\phi_0)$ is called generating functional since $\phi_0$
becomes a source term which couples to a composite operator in the
dual $CFT$.  Another partition function $Z^\prime$ was
constructed\cite{Petkou1} from $Z_{hol}$,
\begin{equation}
Z^\prime=\int[D\phi_0]e^{W(\phi_0)}Z_{hol}, 
\end{equation}
which is a partition function with a new non-trivial weight, $e^{W(\phi_0)}$.
On the other hand, stochastic partition function is spelled out as
\begin{equation}
Z_{SQ}=\int [D\phi_0]e^{-\frac{S_c(\phi_0)}{2}}[D \phi]e^{-S_{FP}(\phi)},
\end{equation}
where $S_{FP}$ is called Fokker-Planck action and $S_c(\phi_0)$ is
classical action which appears in the boundary at $t=0$, where the
stochastic time evolution starts from $-\infty$ and ends up with
$t=0$, i.e. $-\infty<t<0$.  Fokker-Planck action can be made out of
the classical action by promoting the boundary field $\phi_0
\rightarrow \phi_0(t)$ \footnote{For a detailed discussion, see
  Sec.\ref{Stochastic quantization and holographic Wilsonian
    renormalization group}.}.

After identifying the two different partition functions,
$Z^\prime$ and $Z_{SQ}$, it was conjectured\cite{Petkou1} that there
is a one to one correspondence between the Fokker-Planck action
$S_{FP}$ and the classical action $S_c$ in stochastic partition
function, and the bulk action $S_{bulk}$ and the generating
functional $W$ in holographic partition function respectively,
provided that the stochastic time `$-t$' is identified to the
radial coordinate `$r$'.

In fact, in \cite{Sebastian2,Sebastian3,de Haro:2006nv,deHaro:2006wy}, the authors have
studied conformally coupled scalar field theory in $AdS_4$ and
obtained a boundary on-shell action at the conformal boundary. The
boundary action becomes scalar field theory action with two derivative
kinetic term and 6-point self interacting vertex by truncation up to
leading order in large conformal coupling expansion.
It follows from their proposal that $S_c=-2 W(\phi)$, and one can then
construct the Fokker-Planck action and which reproduces the bulk
action (again) by truncation up to leading order in large coupling
expansion of boundary $\phi^6$ interaction.

These two independent results motivated us to study relation between
the holographic renormalization group and the stochastic quantization.
The main motivation is that if such an identification can reproduce
the Fokker-Planck action using boundary on-shell action, then one may
be able to reconstruct radial evolution of the boundary effective
action via stochastic time evolution using this Fokker-Planck action.

In this paper, we have developed a one to one correspondence between
these two schemes, the Holographic Wilsonian Renormalization Group and
the Stochastic quantization, by analyzing their Hamiltonian formalism for
scalar fields and abelian gauge fields such that their dynamics in the
$AdS_4$ space is reproduced in the limit of two derivative bulk
actions.  While the Holographic Wilsonian Renormalization Group is
closely tied with the $AdS$ geometry, the Hamiltonian formalism for
Stochastic processes has no a priori relation with AdS/CFT.  As will be
explained in Sec.\ref{Stochastic quantization and holographic
  Wilsonian renormalization group}, the Hamiltonian formalism is
suitable for both of holographic renormalization group and stochastic
quantization.  For holographic renormalization group, it is given by
\begin{equation}
\partial_\epsilon \psi_H(\phi,r)=-
\int_{r=\epsilon}d^dx \mathcal H_{RG}(-\frac{\delta }{\delta \phi},\phi)\psi_H(\phi,r),
\end{equation}
where $\mathcal H_{RG}$ is obtained from the bulk action by Legendre
transform and $\psi_H=e^{-S_B}$, where $S_B$ is boundary
deformation(boundary effective action).  The stochastic Hamiltonian
formalism, on the other hand gives
\begin{equation}
\partial_t \psi_S(\phi,t)=-\int d^d x \mathcal H_{FP}(\frac{\delta}{\delta \phi},\phi)\psi_S(\phi,t),
\end{equation}
where $\mathcal H_{FP}$ is called the Fokker-Planck Hamiltonian, which
is related to the Fokker-Planck action by Legendre transform.
The wave function $\psi_S(\phi,t)$ is given by
\begin{equation}
\psi_S(\phi,t)= P(\phi,t)e^{\frac{S_c(\phi(t))}{2}},
\end{equation}
where $P(\phi,t)$ is called the probability distribution, which
provide non-trivial weight for the stochastic partition function.

We focus on the similarity between them, and developed one to one
correspondence of quantities appearing in each Hamiltonian
dynamics. We briefly discuss our proposal here.  It is easy to note by
comparing these two Hamiltonian dynamics that, {\bf(1) the stochastic
  time $t$ should be identified to the radial coordinate $r$}, which
is a statement similar to that in\cite{Petkou1}, but this time, {\bf
  precisely $r=t$}.  We also assume that in the absence of any deformation
terms at the boundary of the AdS space {\bf(2) the classical action,
  $S_c$ and the on-shell action, $I_{os}(\phi)$ (or classical
  effective action, $\Gamma(\phi)$ through Legendre transform from it)
  are related as $S_c\equiv2\Gamma(\phi)=-2I_{os}(\phi)$.}  Finally,
we identify these two different Hamiltonians as {\bf (3) $\mathcal
  H_{RG}(r)=\mathcal H_{FP}(t)$}, which is consistent with proposal
(1).

We will discuss our proposal in detail in Sec.\ref{Relations between
  Stochastic quantization and Holographic Wilsonian renormalization
  group}.  Here we would like to summarize the reason for proposing
the
identification(2). As mentioned in the discussion of the holographic
renormalization in pure $AdS$, there is a single $IR$ fixed point for
most of the curves(flows) and the $IR$ effective action has the same
form as the classical effective action $\Gamma$ on the conformal
boundary.  Similar phenomenon happens in the case of stochastic
quantization.  One starts with a system described by a classical
action $S_c$.  Under stochastic time evolution the system will settle
in an equilibrium state which can be described in terms of the
Euclidean partition function with an action which provides
quantization of $S_c$ at some late time $t$.  Therefore, if we impose
identification (2) then we are, at least, guaranteed that most of the
$IR$ behavior of the holographic renormalization group flow and the
late time behavior of stochastic time evolution are the same. $UV$
behavior, as we will see, turns out to be dependent on the initial
condition for the stochastic time evolution.  We will discuss
appropriate choice of initial condition in Sec.\ref{Examples}.

Another point that we would like to mention here is related to the
conjecture (3). This is a non-trivial statement since the form of
Fokker-Planck Hamiltonian density is completely determined by $S_c$. Therefore,
conjecture (3) completely depends on the proposal (2) and it could be
a conditional statement. However, we believe that if certain specific
choice of $S_c$ gives rise to correct $IR$ behavior(equivalently, late
time behavior), then that $S_c$ will provide a correct(similar for the
weak condition) form of the Fokker-Planck Hamiltonian.

We have obtained the following relations as a consequence of our
proposal. Firstly, we have found that double trace deformation part of the boundary effective action,
{\bf (1) $S_B$ is given by}
\begin{equation}
S_B= \int^t_{t_0} d\tilde t \int d^dx \mathcal L _{FP}(\phi(t,x)),
\end{equation}
 in the classical limit, which is the main result of this paper.  Secondly, we have studied the
Langevin dynamics to establish {\bf (2) the relation between
  stochastic 2-point correlation functions and the double trace
  coupling in AdS/CFT}
\begin{equation}
<\phi_{q}(t)\phi_{q^\prime}(t)>^{-1}_H=<\phi_{q}(t)\phi_{q^\prime}(t)>^{-1}_S-\frac{1}{2}\frac{\delta^2
  S_c}{\delta \phi_q(t) \delta \phi_{q^
\prime}(t)},
\end{equation}
where 
\begin{equation}
<\phi_{q}(t)\phi_{q^\prime}(t)>^{-1}_H=
\frac{\delta^2 S_B}{\delta \phi_q(t) \delta \phi_{q^\prime}(t)},
\end{equation}
the coefficient of double trace deformation term and
$<\phi_{q}(t)\phi_{q^\prime}(t)>_S$ is stochastic 2-point correlation
function.

To test our proposal, we have worked out two examples, which are
(minimally coupled)massless scalar fields in $AdS_2$ and $U(1)$ gauge
field theory in $AdS_4$.  It turns out that stochastic quantization
successfully captures the radial evolution of double trace couplings
appearing holographic renormalization group computations of these
examples through the above two relations.

These two models presents several interesting features. Firstly, they
allow alternative quantization. Secondly, their actions are Weyl
invariant. The first condition provides a good playground for
analyzing a variety of boundary conditions, which means the model will
provide more than one fixed point on the $UV$ boundary and diverse
radial flows. The second condition will make computations easy because
Weyl invariance implies there will be no divergent behavior of the
bulk modes and as a result the counter-term action is not necessary.
Another merit of the second condition is that bulk action will
effectively defined on the flat space(See beginning of
Sec.\ref{Examples} for details).

Finally, it turns out that the Fokker-Planck Hamiltonian(Lagrangian)
density obtained from such a classical action, $S_c=2\Gamma$
approximately reconstructs the form of the bulk Hamiltonian(Lagrangian)
density, therefore conjecture (3) is partially proved in these cases.
For the massless scalar field case, the bulk Lagrangian is completely
reconstructed. However, The $U(1)$ gauge fields case is not since to
evaluate boundary on-shell action we have chosen a gauge. Therefore,
the bulk Lagrangian is recovered up to gauge degrees of freedom.
%


\section{$AdS/$(free)$CFT$ and Stochastic Quantization}
\setcounter{equation}{0}
\label{adscft and stochastic quantization}
\subsection{Stochastic Quantization and Holographic Wilsonian Renormalization Group}
\label{Stochastic quantization and holographic Wilsonian renormalization group}
In this section, we will discuss similarity between holographic
Wilsonian renormalization group flows($HWRG$)\cite{Polchinski1,Hong1}
and stochastic quantization($SQ$)\cite{Huffel1,Dijkgraaf1}.  We will
set up one to one mapping between various quantities such as the two-point
correlators, boundary effective actions and so on, appearing in the $HWRG$
and those in the $SQ$.
\subsubsection{Holographic Wilsonian Renormalization Group}
In this subsection, we briefly review the {\it HWRG}.
We start with a bulk action in the Euclidean $AdS_{d+1}$ as
\begin{equation}
\label{total-action}
S=\int_{r>\epsilon} drd^{d}x \sqrt{g}\mathcal  L(\phi,\partial\phi) +  S_{B},
\end{equation}
where $\epsilon$ is an arbitrary cut-off in the radial direction.
The background $AdS$ metric is given by
\begin{equation}
ds^2=\frac{dr^2+\sum^{d}_{i=1}dx_i dx_i}{r^2},
\end{equation}
and $S_B$ is interpreted as the boundary effective action.

{}From the condition that variation of the full action $S$
vanishes, one can define the canonical momentum $\Pi_\phi$:
\begin{equation}
\Pi_\phi=\sqrt{g}\frac{\partial \mathcal L}{\partial(\partial_r
  \phi)} = \frac{\delta S_B}{\delta \phi(x)},
\end{equation}
as a boundary condition.  Since the cut-off $\epsilon$ in the
action(\ref{total-action}) is arbitrary, the physical requirement that the total
action(\ref{total-action}) does not depend on the cut-off $\epsilon$
gives rise to the following equation:
\begin{equation}
\label{hrg-flow-equation}
\partial_\epsilon S_B=-\int_{r=\epsilon}d^dx\left(\frac{\delta S_B}{\delta \phi} 
\partial_r \phi-\mathcal L(\phi,\partial \phi) \right)
=\int_{r=\epsilon}d^dx \mathcal H_{RG}(\frac{\delta S_B}{\delta \phi},\phi),
\end{equation}
where for the second equality in (\ref{hrg-flow-equation}), we have
performed Legendre transform from the Lagrangian density, $\mathcal
L$, using the definition of canonical momentum $\Pi_\phi$ to $\mathcal
H_{RG}$ which is the Hamiltonian density.  The
eq.(\ref{hrg-flow-equation}) is, in fact, semi-classical version of
the Schr\" odinger type equation. To see this more
precisely, one can define the wave functional $\psi$ as
\begin{equation}
\psi_H=\exp(-S_B),
\end{equation}
and the Schr\" odinger type wave equation is 
\begin{equation}
\label{Hamiltonian-wave-equation}
\partial_\epsilon \psi_H=-
\int_{r=\epsilon}d^dx \mathcal H_{RG}(-\frac{\delta }{\delta \phi},\phi)\psi_H.
\end{equation}
In this discussion,
we have implicitly assumed that the Hamiltonian density is quadratic
in canonical momentum.  Eq.(\ref{hrg-flow-equation}) is recovered in
the semi-classical limit, {\it i.e.}, $\left(\frac{\delta S_B }{\delta
    \phi}\right)^2 >> \frac{\delta^2 S_B }{\delta \phi^2}$ and
ignoring terms proportional to $\frac{\delta^2 S_B }{\delta \phi^2}$.

\subsubsection{Stochastic Quantization}
The Hamiltonian description of a system in terms of fictitious
stochastic time `$t$' is defined in the stochastic
quantization\footnote{For reviews, see \cite{Huffel1,Dijkgraaf1}.} as
well.
We will now briefly discuss the method of stochastic quantization, for
which we mostly follow \cite{Huffel1}. The basic notion of stochastic
quantization comes from the similarity between partition function of
Euclidean field theory and partition function of a statistical system in
equilibrium.  The Euclidean $N$-point correlation function is given by
\begin{equation}
\label{stochastic-euclidean-correlations}
<\phi(x_1)...\phi(x_N)>=\int
D\phi\frac{e^{-\frac{1}{\hbar}S_c(\phi)}}{\int D\tilde\phi
  e^{-\frac{1}{\hbar}S_c(\tilde \phi)}} \phi(x_1)...\phi(x_N), 
\end{equation}
where $S_c$ is an Euclidean action(It is also called the `classical
action').  However, once we identify $\hbar \equiv k_{B}T$, where
$k_B$ is Boltzmann constant and $T$ is temperature, this partition
function can also be interpreted as the partition function of a
statistical system in equilibrium with a bath at temperature $T$.
Stochastic process describes evolution of a statistical system from a
non-equilibrium configuration, along a fictitious time to an
equilibrium configuration at the very late time. The fictitious time
here is called the stochastic time 
and it is different from Euclidean time $x_0 \equiv \tau$. Unlike in
the equilibrium state, the measure in
eq.(\ref{stochastic-euclidean-correlations}) for 
non-equilibrium states is not a Boltzmann distribution. Therefore, we
define correlation functions in non-equilibrium states with a general
measure $P(\phi,t)$(which is called the probability distribution)as
\begin{equation}
\label{general-correlations}
<\phi(x_1)...\phi(x_N)>=\int D\phi P(\phi,t) \phi(x_1)...\phi(x_N).
\end{equation}
Technically, stochastic process is describing stochastic time
evolution of $P(\phi,t)$ and once $P(\phi,t)$ is known, then the
correlation functions during stochastic precess are entirely known.
\paragraph{The Langevin Dynamics}
The first realization of this idea was given by Parisi and
Wu\cite{Wu1}. To understand their treatment, let us consider $\phi(x)$
which is a scalar field in $d$-dimensional space with a classical
action, $S_c$.  We suppose that this field $\phi(x)$ interacts with an
imaginary thermal reservoir with temperature $T$ and the system
evolves, by interacting with this thermal reservoir, along the
fictitious stochastic time $t$. Since the system is evolving with time
$t$, we promote the field $\phi(x)$, for it to be time dependent, to
\begin{equation}
\phi(x)\rightarrow \phi(x,t)
\end{equation}
and we expect that in large $t$ limit the system approaches a state of
thermal equilibrium state.

It turns out that the relaxation process satisfies the following equation of motion:
\begin{equation}
\label{langevin-eq}
\frac{\partial \phi(x,t)}{\partial t}=-\frac{1}{2}\frac{\delta S_c}{\delta \phi(x,t)}+\eta(x,t),
\end{equation}
which is called the Langevin equation,
where $\eta(x,t)$ is the Gaussian white noise, which provides
interactions with thermal reservoir.  This white noise has Gaussian probability distribution and
its expectation values are defined as
\begin{equation}
<\eta(x_1,t_1)...\eta(x_N,t_N)>=\frac{\int D\eta(x,t)\eta(x_1,t_1)...\eta(x_N,t_N)e^{-\frac{1}{2}\int d^dxdt \eta^2(x,t)}}{\int D\eta(x,t)e^{-\frac{1}{2}\int d^dxdt \eta^2(x,t)}}.
\end{equation}
Explicit computations of these correlation functions provide rules for
the correlations of $\eta(x,t)$ namely
\begin{eqnarray}
\label{commutation relation of eta}
<\eta_{i,q}(t)>&=&0 {\ , \ } <\eta_{i,q}(t)\eta_{j,q^{\prime}}(t^{\prime})>=\delta_{ij}\delta^d(q-q^\prime)\delta(t-t^\prime),\\ \nonumber
<\eta_{i_1,q_1}(t_1)...\eta_{i_{2k},q_{2k}}(t_{2k})>&=&\sum_{\rm\ all\ possible\ pairs\ of\ i\ and\  j}\Pi_{\rm pairs}<\eta_{i,q_i}(t_i)\eta_{j,q_{j}}(t_{j})>,
\end{eqnarray}
and any correlations with odd number of insertions of $\eta_{i,q}(t)$
vanish. 

Finally, to obtain correlation functions of $\phi(x,t)$, we need to
solve the Langevin equation and get solution of $\phi(x,t)$ with
explicit dependence on $\eta(x,t)$, then we get
\begin{equation}
<\phi(x_1,t_1)...\phi(x_N,t_N)>
=\frac{\int D\eta(x,t)\phi(x_1,t_1)...\phi(x_N,t_N)e^{-\frac{1}{2}\int d^dxdt \eta^2(x,t)}}{\int D\eta(x,t)e^{-\frac{1}{2}\int d^dxdt \eta^2(x,t)}}. 
\end{equation}

\paragraph{Obtaining the probability distribution from the Langevin dynamics}
As we mentioned, getting probability distribution $P(\phi,t)$ is very
crucial for stochastic process. Let us get into the details for this.
The partition function for Langevin dynamics is
\begin{equation}
Z=\int D\eta(x,t)e^{-\frac{1}{2}\int d^dxdt \eta^2(x,t)}.	
\end{equation}
To get more useful information from the partition function, it is
convenient to switch from $\eta(x,t)$ to $\phi(x,t)$ in the partition
function by using the Langevin equation(\ref{langevin-eq}),
\begin{equation}
\label{primitive-partition-f}
Z=\int D\phi(x,t)det\left( \frac{\delta \eta}{\delta \phi} \right)
P(\phi,t_0)exp\left[{-\frac{1}{2}\int^t_{t_0} d^d x d \tilde t\left(
      \dot \phi(x,\tilde t)  
      +\frac{1}{2} \frac{\delta S_c}{\delta \phi(x,\tilde t)}\right)^2}\right],
\end{equation}
where
\begin{equation}
P(\phi,t_0)=\Pi_x \delta^d\left( \phi(x,t_0)-\phi_0(x) \right),
\end{equation}
which gives the initial condition for $\phi(x)$, $t_0$ is initial time
and `dot' denotes derivative with respect to $\tilde t$.
The Jacobian factor can be written more explicitly using the Langevin
equation as,
\begin{equation}
det\left( \frac{\delta \eta}{\delta \phi}  \right)=\exp\left[
  \frac{1}{4}\int^t_{t_0} d\tilde t \int d^dx \frac{\delta^2
    S_c}{\delta \phi^2(x,\tilde t)} \right]. 
\end{equation}
Once we expand the exponent of (\ref{primitive-partition-f}), it gives
a total derivative term with respect to $\tilde t$. 
This total derivative term provides
boundary contribution at $\tilde t=t_0$ and $\tilde t=t$.  With all
this taken into account we get
\begin{equation}
Z=\int
D\phi(x,t_0)P(\phi,t_0)e^{\frac{S_c(\phi(t_0))}{2}}D\phi(x,t)e^{-\frac{S_c(\phi(t))}{2}}[D\phi]\exp\left(
  -\int^{t}_{t_0} d\tilde t \int d^dx \mathcal L
  _{FP}(\phi(\tilde t,x))\right), 
\end{equation}
where 
\begin{equation}
[D\phi]=\Pi_{t_0<\tilde t<t}D\phi(x,\tilde t)
\end{equation}
and 
\begin{equation}
\label{Fokker-Planck-Lagrangian}
\mathcal L_{FP}=\frac{1}{2}\left(\frac{\partial\phi(x)}{\partial t}\right)^2+\frac{1}{8}\left(\frac{\delta S_c}{\delta \phi(x)}  \right)^2-\frac{1}{4}\frac{\delta^2 S_c}{\delta \phi^2(x)},
\end{equation}
which is called the Fokker-Planck Lagrangian density. From this expression, N-point correlation functions can be easily computed. By comparison this with (\ref{general-correlations}), 
one can write the probability distribution function as
\begin{equation}
\label{Useful-exp-probability-distribution}
P(\phi,t)=\exp\left[  -\frac{S_c(\phi(t))}{2} -\int^t_{t_0} d\tilde t
  \int d^dx \mathcal L _{FP}(\phi(\tilde t,x)) \right]. 
\end{equation}

\paragraph{The Fokker-Planck Approach}
The equation satisfied by the probability distribution $P(\phi)$ can
be derived using the Langevin equation,
\begin{equation}
\frac{\partial{P(\phi,t)}}{\partial t}=\frac12\int d^d x \frac{\delta}{\delta \phi(x,t)}\left( \frac{\delta S_c}{\delta \phi(x,t)}+\frac{\delta}{ \delta \phi(x,t)} \right) P(\phi,t).
\end{equation}
We will express this equation in a more suggestive form by defining a wave function $\psi_S$ as
\begin{equation}
\psi_S(\phi,t)\equiv P(\phi,t)e^{\frac{S_c}{2}}, 
\end{equation}
and demanding that this wave function satisfies the Schr\" odinger type equation of motion:
\begin{equation}
\label{stochastic-HaMiLtonian-EOM}
\partial_t \psi_S(\phi,t)=-\int d^d x \mathcal H_{FP}(\frac{\delta}{\delta \phi},\phi)\psi_S(\phi,t),
\end{equation}
where $\mathcal H_{FP}$ is called the Fokker-Planck Hamiltonian, which is given by
\begin{eqnarray}
\mathcal H_{FP}&\equiv& \frac{1}{2}\left(-\frac{\delta}{\delta \phi(x)} +\frac{1}{2}\frac{\delta S_c}{\delta \phi(x)}  \right)
\left(\frac{\delta}{\delta \phi(x)} +\frac{1}{2}\frac{\delta S_c}{\delta \phi(x)}  \right) \\ \nonumber
&=&-\frac{1}{2}\frac{\delta^2}{\delta \phi^2(x)}+\frac{1}{8}\left(\frac{\delta S_c}{\delta \phi(x)}  \right)^2-\frac{1}{4}\frac{\delta^2 S_c}{\delta \phi^2(x)}
\end{eqnarray}
In fact, the Fokker-Planck Lagrangian (\ref{Fokker-Planck-Lagrangian}) is related to $\mathcal H_{FP}$ through Legendre transform.

\subsection{Relations between Stochastic Quantization and Holographic
  Wilsonian Renormalization Group}
\label{Relations between Stochastic quantization and Holographic Wilsonian renormalization group}
\paragraph{The Fokker-Planck approach}
In \cite{Petkou1}, it was suggested that some quantities in the 
stochastic quantization may be identified with quantities appearing in
$AdS/CFT$ in the following manner
\begin{itemize}
\item The fictitious stochastic time, `$t$' $\rightarrow$ $AdS$ radial
  coordinate `$r$' from its boundary to the interior,
\item The Fokker-Planck action: $S_{FP}$ $\rightarrow$ The bulk action:
  $S_{bulk}[\phi_I(r)]$,
\item The classical action, $S_{cl}$ $\rightarrow$
  $-2I_{os}[\phi^{(0)}_I]=2\Gamma[\phi^{(0)}_I]$,
\end{itemize}
where $I_{os}$ is the bulk on-shell action, $\Gamma$ is the classical
effective action, the index, $I$, denotes any index that the bulk
fields(we suppress this index in the most of the following
discussion), $\phi_I$ carry and $\phi^{(0)}_I$ denotes the boundary
value of the bulk field on the conformal boundary. In this section, we
will investigate how many of these assumptions are valid and if they are
all valid, then what kind of information in $AdS/CFT$ is
reproduced by using stochastic quantization.  More precisely, we will
figure out a one to one mapping between quantities appearing in the
stochastic quantization and the Holographic Wilsonian renormalization
group.

We start with a comparison between (\ref{Hamiltonian-wave-equation})
and (\ref{stochastic-HaMiLtonian-EOM}).  They look very similar, and
in fact, they will be the same if the following two conditions are
satisfied:
\begin{itemize}
\item Condition 1: Stochastic time `t' is identified to radial
  coordinate `r' in $AdS$ space.
\item Condition 2: The Fokker-Planck Hamiltonian, $\mathcal H_{FP}$
  has the same form(or similar form as a weak condition) as the
  Hamiltonian of holographic renormalization group flow, $\mathcal
  H_{RG}$. The exact relation is given by
\begin{equation}
\mathcal H_{FP}(t) = \mathcal H_{RG}(r){\rm\ \ provided\ \ }r=t.
\end{equation}

\end{itemize}
The condition 1 is similar to the first suggestion of \cite{Petkou1},
listed above. However, the condition 2 is rather non-trivial. It is
hard to see if the two Hamiltonian densities are the same or
not. Since the form of Fokker-Planck Hamiltonian highly depends on the
classical action $S_c$, determination of $S_c$ is therefore very
crucial.  To determine $S_c$, we follow the
suggestion of \cite{Petkou1} namely
\begin{equation}
S_c=2\Gamma(\phi),
\end{equation}
and we demand that {\it this form of classical action reproduces the
  same(or similar for a mild condition) relation between $\mathcal
  H_{RG}$ and the Fokker-Planck Hamiltonian $\mathcal H_{FP}$.}  If
this condition is satisfied then the Fokker-Planck Lagrangian density
can be derived by Legendre transform and the second condition from
\cite{Petkou1} will also be satisfied.  We therefore propose that
$S_c=2\Gamma(\phi)$ gives the correct choice of the classical action
$S_c$.

Under these conditions, the two Hamiltonian equations of motion (the
Fokker-Planck and the Renormalization Group) are identified.  As a
consequence of this, the two wave functions $\psi_H$ and $\psi_S$ will
also be identified as
\begin{equation}
\label{main-main}
\psi_H=e^{-S_B} \equiv \psi_S=P(\phi,t)e^{\frac{S_c}{2}}. 
\end{equation}
In the classical limit, by using the expression of the probability
distribution (\ref{Useful-exp-probability-distribution}), we can write
$S_B$ explicitly in terms of $\mathcal L _{FP}$ as
\begin{equation}
\label{brief-main}
S_B= \int^t_{t_0} d\tilde t \int d^dx \mathcal L _{FP}(\phi(\tilde t,x)).
\end{equation}
In the limit of $t \rightarrow \infty$(the same with $r \rightarrow
\infty$), $P(\phi,t=\infty)$ is expected to become the Boltzmann distribution
and in that limit, $S_B$ will become
\begin{equation}
\label{main-result}
e^{-S_B} =e^{-S_c + \frac{S_c}{2}}= e^{-\frac{S_c}{2}}=e^{-\Gamma(\phi)}.
\end{equation}
Therefore, at the very late time, $S_B$ converges to $\Gamma(\phi)$,
which is consistent with the $IR$ effective action from the
Holographic Wilsonian renormalization group flows.
\paragraph{Langevin approach}
Equation (\ref{main-main}) also provides a relation between
deformation couplings in the holographic effective action and
correlation functions in stochastic quantization.  For a simple case,
we assume that the theory that we are dealing with is a free theory, so only
two point correlators are non-trivial.  From the definition of
stochastic correlations(\ref{general-correlations}), the two point function
is given by
\begin{equation}
 <\phi_{q_1}(t_1)\phi_{q_2}(t_2)>_S=\int D\phi e^{-S_P(t)} \phi_{q_1}(t_1)\phi_{q_2}(t_2),
\end{equation}
where we define a new quantity $S_p$ as $P(\phi,t) \equiv
e^{-S_P(t)}$.  Since we have assumed that this is a free theory,
$S_P(t)$ will have the form
\begin{equation}
 S_P(t)=\frac{1}{2}\int \mathcal K_q(t)\phi_q(t)\phi_{-q}(t)d^dq,
\end{equation}
where $\mathcal K_q(t)$ is the kernel and $q$ is the d-dimensional
momentum.  From this definition, the two point (equal time)
correlation function in stochastic quantization is
\begin{equation}
\label{stocahstic-corr}
<\phi_{q_1}(t)\phi_{q_2}(t)>_S=\frac{1}{\mathcal K_q(t)} \delta^d(q_1+q_2).
\end{equation}
Notice that in AdS/CFT, double trace couplings in holographic
effective action have a slightly different definition.
According to the relation(\ref{main-main}), $S_B=S_P-\frac{S_c}{2}$ in
the limit of free theory, we define a kernel of the double trace
operator in
holographic effective action as
\begin{equation}
<\phi_{q}(r)\phi_{q^\prime}(r)>^{-1}_H=\frac{\delta^2 S_B} {\delta \phi_{q}(r)\delta \phi_{q^\prime}(r)}.
\end{equation}
{}From the relation (\ref{main-main}), we have
\begin{equation}
\label{holographi-corr}
<\phi_{q}(r)\phi_{q^\prime}(r)>_H=\frac{1}{\mathcal K_q(r)-\tilde k_q(r)} \delta^d(q+q^\prime).
\end{equation}
$\tilde k_q(r)$ is the kernel of $S_c$, we have defined $S_c$ as
\begin{equation}
S_c=\int \tilde k_q(r) \phi_q(r) \phi_{-q}(r)d^dq,
\end{equation}
and the kernel $\tilde k_q(r)$ is formally given by,
\begin{equation}
\label{SC-kelnel}
k_{q}(r)\delta^d(q+q^\prime)=\frac{1}{2}\frac{\delta^2 S_c}{\delta \phi_{q}(r) \delta \phi_{q^\prime}(r)}.
\end{equation}

Therefore, by comparing (\ref{stocahstic-corr}) with
(\ref{holographi-corr}) and using (\ref{SC-kelnel}), we conclude that
there is a relation between two point correlators on both sides as
\begin{equation}
\label{Langevin-finally}
<\phi_{q_1}(t)\phi_{q_2}(t)>^{-1}_H=<\phi_{q_1}(t)\phi_{q_2}(t)>^{-1}_S-\frac{1}{2}\frac{\delta^2 S_c}{\delta \phi_q(t) \delta \phi_{-q}(t)},
\end{equation}
where we identify $r$ to $t$ and $\delta$-function in the momentum
space is ignored in this relation. 

\section{Examples}
\label{Examples}
\setcounter{equation}{0}
\subsection{The simplest example, massless scalar fields in $AdS_2$}
We start with a very simple model, (minimally coupled)massless scalar
field(or zero form field) in Euclidean $AdS_2$. The action is given by
\begin{equation}
\label{scalar-bulk-action}
S_{bulk}=\frac{1}{2}\int dr d\tau \sqrt{g} g^{\mu\nu}\partial_\mu \phi \partial_\nu \phi,
\end{equation}
where $g_{\mu\nu}$ is $AdS_2$ metric and $g$ is its
determinant. $AdS_2$ metric is given by
\begin{equation}
ds^2=\frac{dr^2+d\tau^2}{r^2},
\end{equation}
where $r$ is radial coordinate in $AdS$ with $0 \leq r \leq \infty$,
$r=0$ is $AdS$ boundary and $r=\infty$ is Poincar\' e horizon.  $\tau$ is
Euclidean time(We reserve $t$ to denote the stochastic time.).

Although this is very simple example, it has several merits. First of
all, this action is Weyl invariant.  The Weyl invariance is manifest
since the background metric has a form of
\begin{equation}
g_{\mu\nu}=\frac{\delta_{\mu\nu}}{r^2},
\end{equation}
and substituting this metric into the
action(\ref{scalar-bulk-action}), we get
\begin{equation}
\label{flat-R4-action-scalar}
S_{bulk}=\frac{1}{2}\int_{\mathbb R^2_+} dr d\tau \partial_\mu \phi \partial_\mu \phi,
\end{equation}
where the space-time indices are contracted by $\delta^{\mu\nu}$,
which is the Kronecker $\delta$ and $\mathbb R^2_+$ denotes, say, the
space corresponding to the upper half of $\mathbb R^2$, since the
coordinate `$r$' is semi-infinite. Another feature is that due to this
Weyl invariance, there are no divergent terms in the bulk action as
$r\rightarrow 0$. Therefore, no counter term action is necessary.

Secondly, this action allows `alternative quantization' for its
boundary $CFT$. It is well-known that in $AdS/CFT$, for a particular
range of mass square of bulk scalar fields, $-\frac{d^2}{4} \leq m^2
\leq-\frac{d^2}{4}+1$, there are two possible quantizations. Here $d$
is dimension of boundary space-time.  Each quantization scheme depends
on the boundary condition of the bulk field, which is either Dirichlet
or Neumann boundary condition.  For $AdS_2 $ case, $d=1$ and the mass
square range is given by $-\frac{1}{4} \leq m^2 \leq
\frac{3}{4}$. Therefore, massless scalar fields admits `alternative
quantization'.

\subsubsection{Bulk Solutions and their Boundary On-shell Actions}
\label{Bulk solutions and their boundary effective actions}
\paragraph{Holographic Boundary On-shell Action}
In this section, we apply standard $AdS/CFT$ techniques to our model
and find out its boundary on-shell action.  To obtain this, we will
solve bulk system in the limit of Einstein gravity. In a given $AdS$
background, we get an equation of motion of the scalar field as
\begin{equation}
\label{scalar-eom}
0=(\partial^2_r+\partial^2_\tau)\phi(r,\tau).
\end{equation}
We will solve this equation in the momentum space by using Fourier transform,
\begin{equation}
\label{Fourier transp}
\phi(r,\tau)=\frac{1}{\sqrt{2\pi}}\int^{\infty}_{-\infty} {d\omega} \phi_\omega(r)e^{-i\omega \tau}.
\end{equation}
Then, the equation of motion(\ref{scalar-eom}) becomes
\begin{equation}
\label{scalar-eom-f}
0=(\partial^2_r-\omega^2)\phi_\omega(r).
\end{equation}
The most general form of the bulk solution is given by
\begin{equation}
\label{the-general-scalar-solution}
\phi_\omega(r)=\phi^{(0)}_\omega
\cosh(|\omega|r)+\frac{\phi^{(1)}_\omega}{|\omega|}\sinh(|\omega|r), 
\end{equation}
where $\phi^{(0)}_\omega$ and $\phi^{(1)}_\omega$ are arbitrary
frequency dependent functions.
Another condition that we need to consider is the regularity of the
solution on the Poincar\' e horizon.  The
solution(\ref{the-general-scalar-solution}) is exponentially growing
as in the interior and is divergent at $r=\infty$. To
prevent such a behavior, we set
\begin{equation}
\phi^{(0)}+\frac{\phi^{(1)}}{|\omega|}=0.
\end{equation}
Then, the solution becomes
\begin{equation}
\phi_\omega(r)=\phi^{(0)}_\omega e^{-|\omega|r}.
\end{equation}

Substituting this solution back in the action, up to equation of
motion, we get
\begin{equation}
\label{on-shell action of phi!}
S_{bulk}=\lim_{\epsilon \rightarrow 0}{\ }\frac{1}{2}\int_{r=\epsilon} d\omega \phi_\omega(r) \partial_r \phi_{-\omega}(r)
=-\frac{1}{2}\int d\omega |\omega|\phi^{(0)}_\omega\phi^{(0)}_{-\omega},
\end{equation}
where we have used boundary expansion of $\phi_\omega(r)$ as
\begin{equation}
\phi_\omega(r \rightarrow 0)=\phi^{(0)}_\omega-|\omega|\phi^{(0)}_\omega r+ O(r^2).
\end{equation}
On the boundary of $AdS$, we impose the Dirichlet boundary condition,
$\delta \phi=0$. In this case, there is no deformation term need to be
added to the bulk action. Therefore, the bulk action itself
becomes the on-shell action, $I_{os}(\phi)=S_{bulk}(\phi)$.
The boundary value of $\phi(r)$ is then interpreted as a source term
which couples to a composite operator in boundary $CFT$. Thus we can
then identify the on-shell action with the generating functional with
source $\phi$ as $I_{os}(\phi)=W(\phi)$.

To obtain classical effective action $\Gamma$, we define canonical
momentum as
\begin{equation}
\Pi_{\phi,\omega}=\frac{\partial \mathcal L_{bulk}}{\partial \phi^{\prime}_{\omega}}={\phi^{\prime}_{-\omega}(r)}=-{|\omega|\phi_{-\omega}},
\end{equation}
where `prime' denotes derivative with respect to $r$. The classical
effective action is defined by Legendre transform of the generating
functional as
\begin{equation}
\Gamma[\phi]=-\Pi_{\phi,\omega}\phi_{\omega}+W[\phi].
\end{equation}
Then, we get \footnote{In fact, we need to express the classical effective
  action in terms of $\Pi$, which is given by
\begin{equation}
  \Gamma[\Pi]=\frac{1}{2(2\pi)}\int  \frac{d\omega}{|\omega|}\Pi_\omega(\tilde r)\Pi_{-\omega}(\tilde r),
\end{equation}
where $\Pi$ is vacuum expectation value when one imposes Dirichlet
boundary condition.  However, we express this in terms of $\phi$,
since it is more convenient for the later discussion.}
\begin{equation}
\label{classicalgamma-omega}
\Gamma[\phi]=-W[\phi]
=\frac{1}{2}\int d\omega |\omega|\phi^{(0)}_\omega\phi^{(0)}_{-\omega}.
\end{equation}

\paragraph{Zero frequency solution and its boundary on-shell action}
In the limit of $\omega=0$, the bulk equation of motion (\ref{scalar-eom-f}) is given by
\begin{equation}
\partial^2_r \phi=0,
\end{equation}
and its most general solution is
\begin{equation}
\label{zf-sol}
\phi=\phi^{(0)}+\phi^{(1)}r.
\end{equation}
When we impose regularity condition on the solution in the interior(at
$r=\infty$), we are forced to set $\phi^{(1)}=0$.  Now, to get
boundary on-shell action, we substitute (\ref{zf-sol}) into the
expression of on-shell action(\ref{on-shell action of phi!}). This
gives $I_{os}(\phi)=0$, because the regular solution satisfies
$\partial_r \phi=0$. This means that canonical momentum of $\phi$ is
also zero as $\Pi=\partial_r \phi=0$. By Legendre transform, we get
its classical effective action which is zero, $\Gamma(\phi)=0$, too.

\subsubsection{Holographic Wilsonian renormalization group}
We start with (\ref{hrg-flow-equation}) and our two dimensional bulk
Lagrangian(\ref{flat-R4-action-scalar}).  Substitution
(\ref{flat-R4-action-scalar}) into (\ref{hrg-flow-equation}) provides
holographic Hamilton-Jacobi equation as
\begin{equation}
\label{radial-Hamilton-Jacobi}
\partial_\epsilon S_B=-\frac{1}{2}\int_{r=\epsilon}{d\omega}\left( \left( \frac{\delta S_B}{\delta \phi_\omega} \right)
\left( \frac{\delta S_B}{\delta \phi_{-\omega}} \right) -\omega^2 \phi_\omega \phi_{-\omega}\right).
\end{equation}
Let us solve this equation by assumption of the form of $S_B$ as
\begin{equation}
\label{S-B-ansatz}
S_B=\Lambda(\epsilon)+\int \frac{d\omega}{2\pi}\sqrt{\gamma(\epsilon)}\mathcal J(\epsilon,\omega)\phi_{-\omega}-\frac{1}{2}\int\frac{d\omega}{2\pi}\sqrt{\gamma(\epsilon)}\mathcal 
F(\epsilon,\omega)\phi_\omega\phi_{-\omega},
\end{equation}
where $\Lambda(\epsilon)$, $\mathcal J(\epsilon,\omega)$ and $\mathcal
F(\epsilon,\omega)$ are unknown functions of radial cut-off
$\epsilon$, and especially $\mathcal F(\epsilon,\omega)$ is
interpreted as double trace coupling.  $\gamma(\epsilon)$ is
determinant of (one dimensional) induced metric at $r=\epsilon$
hypersurface, in fact, it is given by
$\gamma={\frac{g(\epsilon)}{{g_{rr}(\epsilon)}}}=\frac{1}{\epsilon^2}$.
Putting the ansatz (\ref{S-B-ansatz}) into
(\ref{radial-Hamilton-Jacobi}) and comparing the coefficients of field
$\phi_{\omega}$, we get the following three equations:
\begin{eqnarray}
\partial_\epsilon \Lambda(\epsilon)&=&-\frac{1}{2}\int_{\epsilon}\frac{d\omega}{(2\pi)^2}J(\epsilon,\omega)J(\epsilon,-\omega), \\
\partial_\epsilon J(\epsilon,-\omega)&=&\frac{1}{2\pi}J(\epsilon,\omega)f(\epsilon,-\omega), \\ 
\label{f-equation}
\partial_\epsilon f(\epsilon,\omega)&=&\frac{1}{2\pi}f(\epsilon,-\omega)f(\epsilon,\omega)-2\pi \omega^2,
\end{eqnarray}
where $J(\epsilon,\omega)\equiv\sqrt{\gamma(\epsilon)}\mathcal
J(\epsilon,\omega)$ and
$f(\epsilon,\omega)\equiv\sqrt{\gamma(\epsilon)}\mathcal
F(\epsilon,\omega)$.  We can then plug the definition of
$f(\epsilon,\omega)$ into (\ref{f-equation}) to obtain a equation in
terms of double trace coupling, $\mathcal F$ as
\begin{equation}
\label{f-modified-equation}
r \partial_r \mathcal F(r,\omega)=\mathcal F(r,\omega)+\frac{1}{2\pi}\mathcal F(r,\omega)\mathcal F(r,-\omega)-2\pi\omega^2r^2.
\end{equation}
The Hamiltonian equation of motion of the bulk field $\phi_\omega$
given by
\begin{equation}
\label{hamilton-eq-motion}
\Pi_\omega=\partial_r \phi_{-\omega}, {\rm\ \ and\ \ } \partial_r \Pi_\omega=\omega^2 \phi_{-\omega},
\end{equation}
can be used to seek the solutions of $\Lambda(\epsilon)$,
$J(\epsilon,\omega)$ and $f(\epsilon,\omega)$. They are
\begin{eqnarray}
f(\epsilon,\omega)&=&-2\pi\frac{\Pi_\omega(\epsilon)}{\phi_{-\omega}(\epsilon)}, {\ \ }J(\epsilon,\omega)=-\frac{\beta_\omega}{\phi_\omega(\epsilon)}, \\ \nonumber
{\rm\ \ and\ \ }\partial_\epsilon \Lambda(\epsilon)&=&-\frac{1}{2}\int_{r=\epsilon}
\frac{d\omega}{(2\pi)^2}\frac{\beta_\omega\beta_{-\omega}}{\phi_{\omega}(\epsilon)\phi_{-\omega}(\epsilon)},
\end{eqnarray}
where $\beta_{\omega}$ is an arbitrary frequency dependent function.

\paragraph{Zero frequency solution}
Now, let us evaluate the effective action $S_B$ by using the above
solution. Solution of (\ref{f-equation}) in $\omega=0$ limit is given
by
\begin{equation}
f(r)=-2\pi\frac{\Pi}{\phi}=-\frac{2\pi \chi}{ 1+ \chi r},
\end{equation}
where $\phi$ is a linear combination of independent solutions
\begin{equation}
\phi=A\phi_1+B\phi_2, {\rm\ where\ } \phi_1=1 {\rm\ and\ }\phi_2=r,
\end{equation}
and $A$, $B$ are arbitrary constants and $\chi\equiv
\frac{B}{A}$. Using this solution, we can write the expression for the
double trace coupling $\mathcal F$ as
\begin{equation}
\label{329}
\mathcal F=-\frac{2\pi \chi r}{ 1+ \chi r}.
\end{equation}
It is easy to see that eq.(\ref{329}) has two different fixed points,
$\mathcal F=0$ and $\mathcal F=-2\pi$(These fixed points are solutions
of eq.(\ref{f-modified-equation})).  Another point to note is that, in
the $IR$ region, we have a single fixed point, $\mathcal F=-2\pi$ if
$\chi \neq 0$, therefore almost every flows will end up with that
fixed point.  If $\chi=0$, then in the IR region $\mathcal F=0$ is a
fixed point.  In the $UV$ region also we have these two different
fixed points but their properties are different.  The $\mathcal
F=-2\pi$ is a fixed point if and only if $\chi=\pm\infty$, whereas
$\mathcal F=0$ is a fixed point when $\chi=0$\cite{Hong1}.

{}From the above solution, we can obtain $S_B$ as
\begin{equation}
\label{zero-f-SB}
S_B=\frac{1}{2}\frac{\chi}{ 1+ \chi r}\phi^2,
\end{equation}
where we have evaluated double trace coupling only(We will deal with double trace couplings only in the most of the following discussion.) and integration
over frequency is removed because we are at $\omega=0$(Effectively, we
have inserted $\delta(\omega)$ in the integrand).  In
Sec.\ref{Stochastic quantization of the classical effective action:
  Fokker-Planck approach}, one will see that (\ref{zero-f-SB}) is
precisely reproduced by stochastic quantization.

\paragraph{Solution with non-zero frequency}
The solutions of bulk equations of motion,
(\ref{hamilton-eq-motion}) are linear combination of
$\cosh(|\omega|r)$ and $\sinh(|\omega|r)$ when frequency is turned
on. Using this fact, the effective action $S_B$ is given by
\begin{equation}
\label{effective-HJ}
S_B(r)=\frac{1}{2}\int{d\omega}|\omega|\left(\frac{\sinh(|\omega|r)+\tilde\phi_\omega \cosh(|\omega|r)}{\cosh(|\omega|r)+\tilde\phi_\omega \sinh(|\omega|r)}\right)
\phi_\omega\phi_{-\omega},
\end{equation}
where $\tilde\phi_\omega $ is an frequency dependent real
function\footnote{If $\tilde\phi_\omega $ is not real, then the double trace deformation is not hermitian.}.
As $r\rightarrow\infty$, the boundary action approaches its
$IR$ region in the sense of holographic renormalization group. The
form of $IR$ effective action is
\begin{equation}
S_B(r=\infty)=\frac{1}{2}\int{d\omega}|\omega|\phi_\omega\phi_{-\omega},
\end{equation}
unless $\tilde\phi_\omega=-1$. If $\tilde\phi_\omega=-1$, then 
\begin{equation}
S_B(r)=-\frac{1}{2}\int{d\omega}|\omega|\phi_\omega\phi_{-\omega}.
\end{equation}

\paragraph{$UV$ and $IR$ fixed points of the double trace coupling and its flows}
It is clear that there are several $UV$ fixed points for the double
trace coupling, $\mathcal F$.
In the $UV$-region($r \rightarrow 0$), 
that there are two fixed points as $\mathcal F(r,\omega)=0$ and
$\mathcal F(r,\omega)=-2\pi$, since the last term in
(\ref{f-modified-equation}) vanishes. Classification of these fixed
points depends on boundary conditions, {\it i.e.}, on the choice of
$\tilde\phi_\omega$.  If we choose $\tilde\phi_\omega=\pm \infty$, we
have $\mathcal F(r,\omega)=-2\pi$ fixed point and for $\tilde
\phi_\omega=0$, we have $\mathcal F(r,\omega)=0$ fixed point.
However, it is not certain if $\mathcal F(r,\omega)$ has $IR$ fixed
points from (\ref{f-modified-equation}) since the last term in it
cannot be ignored in large $r$ region anymore.  In fact, from the
solution of double trace coupling,
\begin{equation}
\mathcal F(r,\omega)=-2\pi|\omega|r\frac{\sinh(|\omega|r)+\tilde\phi_\omega \cosh(|\omega|r)}{\cosh(|\omega|r)+\tilde\phi_\omega \sinh(|\omega|r)},
\end{equation}
one can recognize that it converges to a single fixed point, $\mathcal F(r,\omega)=-\infty$ in $IR$ region for any values of $\tilde\phi_\omega$ except
$\tilde \phi_\omega=-1$. If $\tilde \phi_\omega=-1$, $\mathcal F(r,\omega)=\infty$ is $IR$ fixed point.

\subsubsection{Stochastic quantization of the classical effective action: Zero frequency}
\label{Stochastic quantization of the classical effective action: Fokker-Planck approach}
As per our proposal, relation between $S_{cl}$ and $AdS/CFT$ is that
$S_{cl}=2\Gamma[\phi]$, where $\Gamma[\phi]$ is classical effective
action on $AdS_2$ boundary for the massless scalar field.  Leaving out this
connection, we will not use any information from $AdS/CFT$ for our
computations in this section, we will only use stochastic quantization
techniques.

\paragraph{The Fokker-Planck action}
Let us evaluate stochastic time evolution of the system
in which the classical action is given by $S_c=2\Gamma(\phi)$ as
conjectured in Sec.\ref{Relations between Stochastic quantization and
  Holographic Wilsonian renormalization group}.  As we discussed in
Sec.\ref{Bulk solutions and their boundary effective actions}, in the
case of zero frequency, the classical effective action,
$\Gamma(\phi)=0$.  From the expression of Fokker-Planck action, we
have
\begin{equation}
\label{SB-0-frequency}
S_{FP}=\frac{1}{2}\int^t_{t_0}  \left(\frac{\partial \phi}{\partial t}\right)^2 dt.
\end{equation} 
This Fokker-Planck Lagrangian density has precisely the same form as the bulk
action(\ref{flat-R4-action-scalar}) with $\omega=0$ with the
identification, $t=r$.  Let us evaluate $S_{FP}$ in the classical
limit. To do this, we use equation of motion from this action, which
is given by
\begin{equation}
\frac{\partial^2 \phi}{\partial t^2}=0,
\end{equation}
and the most general solution is 
\begin{equation}
\label{G-sol-aa}
\phi=a_1+a_2 t,
\end{equation}
where $a_1$ and $a_2$ are arbitrary real constants.
We will impose boundary conditions to constrain the parameters in
(\ref{G-sol-aa}). Suppose at a certain time $t$, we want to field  $\phi(\tilde t = t)=\phi(t)$,
then, the solution becomes
\footnote{This is the usual boundary condition to evaluate Fokker-Planck action. For example, see Sec.3.2.2. in \cite{Huffel1}}
\begin{equation}
\phi(\tilde t)=\phi(t)\frac{1+a\tilde t}{1+at},
\end{equation}
where, $a=\frac{a_2}{a_1}$. Let us plug this solution into (\ref{SB-0-frequency}), then we get
\begin{equation}
S_{FP}=\left.\frac{1}{2}\phi(\tilde t){\partial_{\tilde t}\phi(\tilde t)}\right|^t_{t_{0}},
\end{equation}
where $t_0$ is initial time.
{\it At this point, we propose that a judicious choice of the initial
  time $t_0$ precisely reproduces holographic renormalization group
  result. The prescription is to set}\footnote{Our prescription for
  the choice of $t_0$ will become clear momentarily when we will
  discuss the Langevin dynamics}
\begin{equation}
t_0=-\frac{1}{a},
\end{equation}
{\it at which the solution (\ref{G-sol-aa}) becomes zero,
  $\phi(t_0=-\frac{1}{a})=0$, and the, the range over which $t$
  varies becomes $-\frac{1}{a}<t<\infty$. Therefore, for the 
  identification of $t=r$, we identify a subset of the interval of
  `$t$' with the interval of `$r$', $0<r<\infty$. 
  When `$a$' is positive, the stochastic process begins before
  $t=0$. In this case, we identify only a subset of the interval of
  $t$ as $0<t<\infty$ to $r$. 
  For the negative value of $a$, the stochastic process begins after
  $t=0$, then we identify entire $-\frac{1}{a}<t<\infty$ with $r$ but
  then it covers only a part of the interval of `$r$'.  Finite non-zero
  value of $r$ corresponds to UV cutoff in AdS/CFT.  Thus for negative
`$a$' , stochastic process gives evolution of a field theory with
explicit UV cutoff.}
With such a choice, we get
\begin{equation}
\label{fokker-planck-zerof}
S_{FP}=\frac{1}{2}\frac{a}{1+at}\phi^2(t),
\end{equation}
which is of the same form as (\ref{zero-f-SB}) once we identify $t$
and $a$ with $r$ and $\chi$ respectively.  We thus see that, in this
case, the prescription (\ref{brief-main}) is correct up to making a
choice of $t_0$.  

\paragraph{Langevin dynamics}
The Langevin equation (\ref{langevin-eq}) in this case $(S_c=0)$ becomes
\begin{equation}
\frac{\partial \phi}{\partial t}=\eta(t),
\end{equation}
where since $\phi$ and $\eta$ do not depend on $\omega$, we demand
\begin{equation}
<\eta(t)\eta(t^\prime)>=\delta(t-t^\prime),
\end{equation}
and $<\eta(t)>=0$.
The solution of Langevin equation is given by
\begin{equation}
\label{phit-solu}
\phi(t)=\int^t_0 \eta(\tilde t)d\tilde t + \phi_0,
\end{equation}
where $\phi_0$ is an integration constant. If $\phi_0$ is chosen
appropriately, then (equal time)two point correlation of $\phi(t)$
will be consistent with holographic RG.

{\it The prescription for choosing $\phi_0$ is}
\begin{equation}
\label{ini-phi0}
\phi_0=\int^0_{-\frac{1}{a}} \eta(\tilde t)d\tilde t.
\end{equation}
{\it We stress that we just choose initial condition for $\phi(t)$ at
  $t=0$. The interval of $t$ is still $0<t<\infty$ for this
  choice. Therefore, 
`$t$' is identified with `$r$'. However, once we plug
(\ref{ini-phi0}) into (\ref{phit-solu}), it has a form}
\begin{equation}
 \phi(t)=\int^t_{-\frac{1}{a}} \eta(\tilde t)d\tilde t.
\end{equation}
{\it Again, $t=-\frac{1}{a}$ is a special point at which the general
  solution(\ref{G-sol-aa}) vanishes, i.e., $\phi(t=-\frac{1}{a})=0$. }
 
With this solution, one can compute (equal time)two point correlation function
\begin{equation}
\label{zero-f-2-point}
<\phi(t)\phi(t)>_S=\int^t_{-\frac{1}{a}}\int^t_{-\frac{1}{a}}<\eta(t^\prime)\eta(t^{\prime\prime})>dt^\prime dt^{\prime\prime}=t+\frac{1}{a}.
\end{equation}
If we identify 
\begin{equation}
\chi=a {\rm\  and \ }r=t,
\end{equation}
then, (\ref{zero-f-2-point}) is precisely matches with
(\ref{zero-f-SB}) through the relation (\ref{Langevin-finally}), since
\begin{equation}
<\phi(r)\phi(r)>_H=\left(\frac{\delta^2 S_B}{\delta \phi(r)\delta \phi(r)}\right)^{-1}=r+\frac{1}{\chi}{\rm \ and\ } \frac{\delta^2 S_c}{\delta \phi(r)\delta \phi(r)}=0.
\end{equation}

\subsubsection{Stochastic quantization of the classical effective action: Non-zero frequency}

\paragraph{Fokker-Planck action}
Our starting point is the classical action obtained from
(\ref{classicalgamma-omega}) using the relation $S_c=2\Gamma(\phi)$
\begin{equation}
\label{classical-classcal-action}
S_{cl}=\int^\infty_{-\infty}d\omega |\omega|\phi_{\omega}\phi_{-\omega}.
\end{equation}

We first evaluate the Fokker-Planck Lagrangian density, which is given by 
\begin{eqnarray}
\mathcal L_{FP}&=&\frac{1}{2}\left( \frac{\partial \phi_\omega}{\partial \tilde t} \right)\left( \frac{\partial \phi_{-\omega}}{\partial \tilde t} \right)
+\frac{1}{8}\left( \frac{\delta S_{cl}}{\partial \phi_\omega} \right)\left( \frac{\delta S_{cl}}{\partial \phi_{-\omega}} \right)
-\frac{1}{4}\left(\frac{\delta^2 S_{cl}}{\delta \phi_{\omega} \delta \phi_{-\omega}}\right) \\ 
\label{FPaction-phi}
&=&\frac{1}{2}\dot \phi_{\omega} \dot \phi_{-\omega}+\frac{1}{2} \omega^2 \phi_{\omega}\phi_{-\omega}-\frac{1}{2}|\omega|\delta(0),
\end{eqnarray}
where to evaluate Fokker-Planck Lagrangian density, we promote the field $\phi_\omega$
\begin{equation}
\phi_\omega \rightarrow \phi_\omega(\tilde t).
\end{equation} 
{\it The Fokker-Planck Lagrangian density has the same form as the bulk
  Lagrangian density (\ref{flat-R4-action-scalar})  
up to a term proportional to the $\delta$-function, which is just an
infinite constant.} This term is not relevant for the following
discussion since it does not 
depend on the field $\phi$.
The `dot' denotes derivative with respect to $\tilde t$, which is the stochastic time. 
We want set the range of the stochastic time as $0 \leq \tilde t \leq
\infty$. 
Therefore, the stochastic system starts from $\tilde t=0$ and settles
in a thermal equilibrium at $\tilde t=\infty$.

Let us evaluate equation of motion, which is given by
\begin{equation}
\label{EOM-ads2-scalar}
0=\ddot \phi_{\omega}-\omega^2 \phi_\omega.
\end{equation}
The most general solution of the equation of motion is
\begin{equation}
\phi_\omega(t)=a_{1,\omega}\cosh(|\omega|t)+a_{2,\omega}\sinh(|\omega|t),
\end{equation}
where $a_{1,\omega}$ and $a_{2,\omega}$ arbitrary frequency dependent functions.

Let us now look at boundary conditions. At certain time $t$, we want
that $\phi_\omega(\tilde t=t)=\phi_\omega(t)$, then, the solution becomes
\begin{equation}
\label{final-solution-phi-pre}
\phi_\omega(\tilde t)=\phi_\omega(t)\frac{\cosh(|\omega|\tilde t)+a_{\omega}\sinh(|\omega|\tilde t)}{\cosh(|\omega|t)+a_{\omega}\sinh(|\omega|t)},
\end{equation}
where $a_\omega=\frac{a_{2,\omega}}{a_{1,\omega}}$.
On substituting the solution (\ref{final-solution-phi-pre}) into
Fokker-Planck action we get,
\begin{eqnarray}
\label{mid-final-form-FP}
S_{FP}&=&\int^{t}_{-\frac{1}{|\omega|}\coth^{-1}(a_\omega)}d\tilde t \int d\omega \left( \frac{1}{2}\dot \phi_{\omega} \dot \phi_{-\omega}
+\frac{1}{2} \omega^2 \phi_{\omega}\phi_{-\omega} \right) =\left.\frac{1}{2}\int d\omega \phi_{\omega}(\tilde t)
 \dot \phi_{-\omega}(\tilde t) \right|^{\tilde t=t}_{\tilde t=-\frac{1}{|\omega|}\coth^{-1}(a_\omega)} \\ \nonumber
&=&\frac{1}{2}\int d\omega |\omega|\phi_{\omega}(t)  \phi_{-\omega}(t)
\left( \frac{\sinh(|\omega| t)+a_{\omega}\cosh(|\omega| t)}{\cosh(|\omega|t)+a_{\omega}\sinh(|\omega|t)}\right),
\end{eqnarray}
where for the second equality, we have used equation of
motion(\ref{EOM-ads2-scalar}) and the lower limit of the integration
is chosen in the same fashion as in Sec.\ref{Stochastic quantization
  of the classical effective action: Fokker-Planck approach}.
The Fokker-Planck action(\ref{mid-final-form-FP}) has the same form as
(\ref{effective-HJ}) with the identification that $a_\omega=\tilde
\phi_\omega$.  Before we discuss the Langevin dynamics let us look at
the reality condition on  $a_\omega$ and $\tilde\phi_\omega$.  The
reality conditions implies
$a^\star_\omega=a_{-\omega}$ and
$\tilde\phi^\star_\omega=\tilde\phi_{-\omega}$ respectively.  In
addition hermiticity of the  Fokker-Planck Lagrangian density 
(\ref{mid-final-form-FP}) implies both $a_\omega$ and
$\tilde\phi_\omega$ are real.

\paragraph{The Langevin dynamics}
Let us now derive the Langevin equation(\ref{langevin-eq}) using
classical action(\ref{classical-classcal-action}), which is given by
\begin{equation}
\frac{\partial \phi_\omega(t)}{\partial t}=-|\omega|\phi_\omega(t)+\eta_\omega(t). 
\end{equation}
The solution with appropriate boundary condition is
\begin{equation}
\phi_\omega(t)=\int^t_{-\frac{1}{|\omega|}\coth^{-1}(a_\omega)}d\tilde t e^{-|\omega|(t-\tilde t)}\eta_\omega(\tilde t).
\end{equation}
Now, we evaluate equal time two point correlator for scalar field using expectation values of
$\eta_{\omega}(\tilde t)$ given in (\ref{commutation relation of eta})
\begin{eqnarray}
\label{langevin-correlatoR}
<\phi_\omega(t)\phi_{\omega^\prime}(t)>_S&=&\int^t_{-\frac{\coth^{-1}(a_\omega)}{|\omega|}}\!\!\!
\!\!\! d \tilde t \int^{t}_{-\frac{\coth^{-1}(a_\omega)}{|\omega|}} 
\!\!\!\!\!\!\!\! d\tilde t^\prime e^{-|\omega|(t-\tilde t)-|\omega^\prime|(t-\tilde t^\prime)}
<\eta_\omega(\tilde t)\eta_{\omega^\prime}(\tilde t^\prime)> \\ \nonumber
&=&\frac{1}{2|\omega|}\left(1-\frac{a_\omega-1}{a_\omega+1}e^{-2|\omega|t}\right)\delta(\omega+\omega^\prime).
\end{eqnarray}
We again see that the eq.(\ref{langevin-correlatoR}) is consistent
with (\ref{effective-HJ}) through the
relation(\ref{Langevin-finally}).

Using the definition of probability
distribution(\ref{Useful-exp-probability-distribution}), we get
\begin{equation}
P(\phi,t)=exp\left[  -\frac{1}{2} \int d\omega|\omega|\left(\frac{(a_\omega+1)e^{|\omega|t}}{\cosh|\omega|t+a_\omega \sinh|\omega|t}\right)  \phi_\omega(t)\phi_{-\omega}(t)\right].
\end{equation}
One can recognize that the kernel in exponent of $P(\phi,t)$ is
precisely the inverse of the two point correlation
$<\phi_\omega(t)\phi_{\omega^\prime}(t^\prime)>_S$ in
(\ref{langevin-correlatoR}).

%
%

At this point we would like to make the following remark. One may suspect that
$t_0=-\frac{1}{|\omega|}\coth^{-1}(a_\omega)$ is not well defined when
$|a_\omega|<1$. 
However, we still assign our boundary condition using this form of
$t_0$, and allowing it to have imaginary part when $|a_\omega|<1$. In fact,
\begin{equation}
t_0=-\frac{1}{|\omega|}\coth^{-1}(a_\omega)=-\frac{1}{|\omega|}\tanh^{-1}(a_\omega)\pm
\frac{i\pi}{2|\omega|} {\rm \ for\ } |a_\omega|<1 .
\end{equation}
To evaluate Fokker-Planck action (\ref{mid-final-form-FP}) in this
case, we choose an integration path in the complex $\tilde t$ plane
and choose positive sign for the imaginary part of $\tilde t$.
The contour is mostly along the real axis except at
$\tilde t=\tanh^{-1}(a_\omega)$ where it goes parallel to imaginary
axis from $\tilde t = \tanh^{-1}(a_\omega)$ to $\tilde t =
\tanh^{-1}(a_\omega) + \frac{i\pi}{2|\omega|}$.  The boundary action
can then be written as
\begin{eqnarray}
\label{SB-smalla}
S_B&=&\int^t_{-\frac{1}{|\omega|}\tanh^{-1}(a_\omega)}d\tilde t\int d\omega \mathcal L_{FP}
+\int^{-\frac{1}{|\omega|}\tanh^{-1}(a_\omega)}_{-\frac{1}{|\omega|}\tanh^{-1}(a_\omega)+\frac{i\pi}{2|\omega|} }d \tilde t\int d\omega \mathcal L_{FP} \\ \nonumber
&=&\left.\frac{1}{2}\int d\omega \phi_{\omega}(\tilde t)
 \dot \phi_{-\omega}(\tilde t) \right|^{\tilde t=t}-\left.\frac{1}{2}\int d\omega \phi_{\omega}(\tilde t)
 \dot \phi_{-\omega}(\tilde t) \right|^{\tilde t=-\frac{1}{|\omega|}\tanh^{-1}(a_\omega)}\\ \nonumber 
&+&\left.\frac{1}{2}\int d\omega \phi_{\omega}(\tilde t)
 \dot \phi_{-\omega}(\tilde t) \right|^{\tilde t=-\frac{1}{|\omega|}\tanh^{-1}(a_\omega)}_{\tilde t=-\frac{1}{|\omega|}\tanh^{-1}(a_\omega)+\frac{i\pi}{2|\omega|}}
\end{eqnarray}
Using the fact that
\begin{equation}
\dot \phi_{\omega}|_{\tilde t=-\frac{1}{|\omega|}\tanh^{-1}(a_\omega)}=0,{\rm \ \ and \ \ }\phi_{\omega}|_{\tilde t=-\frac{1}{|\omega|}\tanh^{-1}(a_\omega)+\frac{i\pi}{2|\omega|}}=0,
\end{equation}
the second  and the third terms in the second equality in (\ref{SB-smalla}) vanishes.
This gives precisely the same result as in (\ref{mid-final-form-FP}). In the first term, the integration variable $\tilde t$ is on the real line, and so is $t$ 
therefore, we will identify $t$ in this case with the $AdS$ radial coordinate $r$.

For the Langevin dynamics, we choose the same integration path
\begin{eqnarray}
\phi_\omega(t)&=&\int^t_{-\frac{1}{|\omega|}\tanh^{-1}(a_\omega)+\frac{i\pi}{2|\omega|}}d\tilde t e^{-|\omega|(t-\tilde t)}\eta_\omega(\tilde t) \\ \nonumber
&=&\int^t_{0}d\tilde t e^{-|\omega|(t-\tilde t)}\eta_\omega(\tilde t)
+\phi_{\omega,0} e^{-|\omega|t},
\end{eqnarray}
where
\begin{equation}
\label{366}
\phi_{\omega,0}=\int^0_{-\frac{1}{|\omega|}\tanh^{-1}(a_\omega)}d\tilde t e^{|\omega|\tilde t}\eta_\omega(\tilde t)
+\int^{-\frac{1}{|\omega|}\tanh^{-1}(a_\omega)}_{-\frac{1}{|\omega|}\tanh^{-1}(a_\omega)
+\frac{i\pi}{2|\omega|}}d\tilde t e^{|\omega|\tilde t}\eta_\omega(\tilde t),
\end{equation}
where $\phi_{\omega}( t=0)=\phi_{\omega,0}$, the initial value of $\phi_{\omega}$.
The integration path in the first integral in (\ref{366}) is a
straight line along the real axis, whereas the path in the second integral
is a straight line parallel to the imaginary axis. 
Again, $t$ is real in the first integral and we identify it with the
$AdS$ radial coordinate $r$.  Let us again check if $\phi_{\omega,0}$
satisfies the reality condition,
$\phi^\star_{\omega,0}=\phi_{-\omega,0}$.  On the real line, it is
sufficient to show this that $\eta^\star_{\omega,0}(\tilde
t)=\eta_{-\omega,0}(\tilde t)$, however, it turns out that the reality
condition will be satisfied along the second contour if we impose another condition
$\eta_{\omega,0}(\tilde t)=\eta_{\omega,0}(\tilde
t+\frac{i\pi}{|\omega|})$ in the complex $\tilde t$ plane.  This is
just a periodicity condition for $\eta_0$ along contour parallel to the imaginary line.


\subsubsection{More on the initial conditions in stochastic quantization}
In the previous discussion for the stochastic quantization, we have
imposed the initial condition as $t=t_0$ in an ad hoc manner.
Here we would like to provide rationale for making such a choice.  For illustration
consider computation of the holographic renormalization group for the zero frequency case. 
The radial flow of the double 
trace operator starts from a fixed point which is either $\chi=0$ or
$|\chi|=\infty$.  Let us, for concreteness, concentrate on the $|\chi|=\infty$ fixed point. 
In this case, the boundary effective
action (\ref{zero-f-SB}) takes the form
\begin{equation}
S_B=\frac{1}{2}\frac{\phi^2}{r}.
\end{equation}
We can also see that the Fokker-Planck action for the zero frequency
case (\ref{fokker-planck-zerof}) can be re-written as
\begin{equation}
S_{FP}=\frac{1}{2}\frac{\phi^2(\hat t)}{\hat t},
\end{equation}
where $\hat t$ is a shifted time coordinate, $\hat t \equiv
t+\frac{1}{a}$ (note that both the Langevin equation and the Fokker-Planck
action possess time translation invariance.).  Comparing the above two
expressions, one realizes that the boundary effective action $S_B$
which starts from $|\chi|=\infty$ fixed point has the same form as the
Fokker-Planck action if we replace $r$
and $\chi$ in the holographic RG by $\hat t$ and $a$
respectively.  This implies that the
choice of the initial time $t_0=-\frac{1}{a}$ in case of the
stochastic process becomes $\hat t_0=0$ and the stochastic system will
start evolving from the fixed point $|a|=\infty$ (which, by our
identification, is equivalent to $|\chi|=\infty$ fixed point in
holographic renormalization group flows). There are several radial
flows of the double trace operator for generic choices of the value of
$\chi$ which do not start from fixed points. However, from the point
of view of the stochastic quantization, all the stochastic time
evolutions begin in the neighborhood of a fixed point but with
different initial times.  At $t=0$ these stochastic time evolutions
are not in the vicinity of any of the fixed points unless $a=0$ or
$|a|=\infty$.  Identification of $r$ is still done with the
stochastic time $t$ only for the range $0\leq r\leq \infty$.  Thus we
see that for all the flows which do not start from the fixed point,
stochastic time evolution starts from $t=-\frac{1}{a}$.  While for
$|\chi|=\infty$ the evolution begins from $t=0$, for $\chi=0_-$ it
begins from $t=-\infty$.
Notice that unlike the radial coordinate of $AdS$ space which cannot
take negative values, the stochastic time can begin with arbitrary
negative values. 

The above scheme for determining initial time is applicable to
non-zero frequency case as well. The final remark is that when
$|a_\omega|<1$, $|\tilde\phi_{\omega}|=\infty$ fixed point will be
obtained by shifting the stochastic time along the imaginary axis as
well as the real axis.  We have therefore chosen a complex initial
time.

\subsection{U(1) gauge fields in $AdS_4$}

We start with the $U(1)$(Euclidean) gauge field action in $AdS_4$
space-time background
\begin{equation}
S_{bulk}[A]=\frac{1}{4}\int d^4x \sqrt{g} F_{\mu\nu}F^{\mu\nu},
\end{equation}
where the space-time indices $\mu,\nu$ run from 1 to 4.  The background metric is
\begin{equation}
\label{AdS4-metric}
ds^2=\frac{dr^2+\delta_{ij}dx^i dx^j}{r^2},
\end{equation}
where the
indices $i,j..$ are defined boundary space-time coordinate, which run
from 1 to 3 or $i=x,y$ and $z$ and $g$ is determinant of metric $g_{\mu\nu}$.  
$U(1)$ field strength is given by
\begin{equation}
F_{\mu\nu}=\partial_{\mu} A_{\nu}-\partial_{\nu} A_{\mu}.
\end{equation}

Like the massless scalar field action in $AdS_2$, this action is also
Weyl invariant and admits alternative quantization.  Under the Weyl
rescaling of background metric, $ds^2 \rightarrow r^2 ds^2$, generic
gauge field theory defined on $AdS_4$ gets mapped to that defined in
4-dimensional flat space-time.  This space-time is only half of $\mathbb
R^4$, because the radial coordinate in $AdS$ space runs from $0$ to
$\infty$. Therefore, the action becomes
\begin{equation}
\label{flat-space-action}
S_{bulk}[A]=\frac{1}{4}\int_{\mathbb R^4_{+}} d^4x  F_{\mu\nu}F^{\mu\nu},
\end{equation}
where, the space-time indices are now contracted with
$\delta_{\mu\nu}$ and $\mathbb R^4_{+}$ denotes a half of the
4-dimensional flat space.

The equations of motion from $S_{bulk}$ are
given by
\begin{eqnarray}
\label{A-bulk-eom}
0&=&\nabla^2 A_{r} -\partial_r \partial_i  A_i, \\ \nonumber
0&=&(\partial^2_r +\nabla^2)A_i-\partial_{i}(\partial_r 
 A_{r}+\partial_j  A_j),
\end{eqnarray}
where
$\nabla^2 \equiv \sum_{j=1}^3 \partial_j \partial_j$.
Solutions to these equations has already been obtained in
\cite{Sebastian1,Jatkar:2012mm}.  Let us briefly recall 
the solution $A_\mu$ in momentum space
\begin{eqnarray}
\label{first order solution}
 A_{i,q}(r)&=&A^{T}_{i,q}(r) -iq_i \phi^a_q(r) , \qquad  \
 A_{r,q}(r)=\partial_r \phi_q(r){, \  \  } q_i 
 A^{T}_{i,q}(r)=0, \\ \nonumber
{\rm \ and \ }  A^{T}_{i,q}(r)&=& A^{T(0)}_{i,q}\cosh(|q|r)
+\frac{1}{|q|} A^{T(1)}_{i,q}\sinh(|q|r),
\end{eqnarray}
where $q_i$ are components of three momentum along the boundary
direction and the solution is obtained by using Fourier transform of
the position space representation defined in a manner similar to
(\ref{Fourier transp}) but this time with three boundary coordinates.
$A^{T}_{i,q}$ is the transverse part of the gauge field, which is given by
\begin{equation}
\bar A^{T}_{i,q} = P_{ij}(q) \bar A_{j,q},
\end{equation}
where we define a projection operator,
\begin{equation}
P_{ij}(q)=\delta_{ij}-\frac{q_i q_j}{q^2},
\end{equation}
and $A^{T(0)}_{i,q}$ and $A^{T(1)}_{i,q}$ are $q_i$ dependent transverse vector functions.
$\phi^a$ is a gauge freedom
which is not completely determined by equations of motion.

To proceed further we will use the radial gauge, namely $
A_{r,q}(r)=0$. In the radial gauge, the residual gauge freedom is
obtained by restricting the gauge parameter $\phi_{r,q}(r)$ to be
independent of $r$,
\begin{equation}
\label{the radial gauge}
\phi_q(r) \rightarrow \phi_q.
\end{equation}
Then by definition, $ A^{T}_{i,q}(r)$ is gauge invariant under this
residual gauge transformation.
Another condition that we need to consider is regularity in the interior of bulk spacetime. 
For the regularity of the solutions at the Poincar\' e horizon, at
$r=\infty$, we require that
\begin{equation}
\label{regularity-condition}
 A^{T(0)}_{ip}+\frac{1}{|p|} A^{T(1)}_{ip}=0.
\end{equation}
This removes the term proportional to $e^{|p|r}$ near the Poincar\' e horizon.
Using this regularity condition we can write the solution in the following form 
\begin{equation}
\label{first order bulk solution}
A^{T}_{i,p}(r)
= A^{T(0)}_{ip}e^{-|p|r}.
\end{equation}
\paragraph{Boundary on-shell action}
Substituting solutions (\ref{first order bulk
  solution}) 
into the bulk on-shell action
\begin{equation}
S_{bulk}[A]=\frac{1}{2}\int d^3q A_{i,q}\partial_r A_{i,-q}.
\end{equation}
we get
\begin{equation}
I_{os}[A]=S_{bulk}[A]=-\frac{1}{2}\int d^3q |q|A^{T(0)}_{i,q} A^{T(0)}_{i,-q},
\end{equation}
which is a manifestly gauge invariant action because it depends only
on the transverse part of the gauge field.  Canonical momentum of the
gauge field is
\begin{equation}
\Pi^T_q=\frac{\delta S_{bulk}[A]}{\delta A^T_{i,q} }=-|q|A^T_{i,-q}.
\end{equation}
The classical effective action $\Gamma[A]$ is then obtained by taking
the Legendre transform of the on-shell action $I_{os}[A]$,
\begin{equation}
\label{gamma-aaa}
\Gamma[A]=-I_{os}[A]=\frac{1}{2}\int d^3q |q|A^{T(0)}_{i,q} A^{T(0)}_{i,-q}.
\end{equation}

\subsubsection{Holographic renormalization group flow of $U(1)$ gauge
  field theory}

We start with the flow equation
\begin{eqnarray}
\partial_\epsilon S_B(A)&=&-\int_{r=\epsilon} \left[ \frac{1}{2} \delta_{ij} \left(\frac{\delta S_B}{\delta A_{i,q}}\right)\left(\frac{\delta S_B}{\delta A_{j,-q}}\right)
-\frac{1}{4}F_{ij,q}F_{kl,-q}\delta_{ik}\delta_{jl}\right]\\ \nonumber
&+&\int (-iq_i)\left( \frac{\delta S_B}{\delta A_{i,q}}\right)A_{r,q}
\end{eqnarray}
in the momentum space.  The holographic renormalization group
computation of $U(1)$ gauge fields is pretty much similar to the
massless scalar field case and is given in \cite{Hong1}. Therefore, we
would like to comment only on the differences between them and then
directly state the result. First of all, the main difference between
them is existence of gauge degrees of freedom, which can be used to
decompose the $U(1)$ gauge fields into transverse and longitudinal
parts.  As argued in \cite{Hong1}, in both cases Dirichlet and
Neumann boundary conditions can be imposed on the conformal boundary,
equations involving transverse components are completely decoupled
from those involving the longitudinal one in holographic Wilsonian RG
computation.  Moreover, we are only interested in radial flows of
double trace coupling of transverse components of the gauge field.
Therefore, the ansatz for $S_B$ is given by
\begin{equation}
S_B(A)=\Lambda(\epsilon)+\int\frac{d^3q}{(2\pi)^3}\sqrt{\gamma}\mathcal J^T_i(q,\epsilon)g^{ij}A^T_{j,q}
-\frac{1}{2}\frac{d^3q}{(2\pi)^3}\sqrt{\gamma}\mathcal F_T(q,\epsilon)g^{ij}A^T_{i,q}A^T_{j,-q},
\end{equation}
where, the superscript(also subscript in some of the later
expressions) `$T$' denotes {\it transverse}.  Again, there are
longitudinal parts in $S_B$, but they are decoupled.

Secondly, the equation and solution of the double trace coupling $\mathcal F_T$ are given by
\footnote{
The other equations are given by
\begin{eqnarray}
\partial_\epsilon \Lambda(\epsilon)&=&-\frac{1}{2(2\pi)^6}\int\delta^{ij}J^T_{i,q}J^T_{j,-q}d^3q, \\ 
\partial_\epsilon J^T_{i,q}&=&\frac{1}{(2\pi)^3}J^T_{i,q}f_{tq},
\end{eqnarray}
where $J^T_{i,q}=\sqrt{\gamma} g^{ij}\mathcal J_{T,j}(q,\epsilon)$.
}
\begin{eqnarray}
\label{equation-RG-A}
\partial_\epsilon f_{i}(q,\epsilon)&=&\frac{1}{(2\pi)^3}f_{i}(q,\epsilon)f_{i}(-q,\epsilon)-(2\pi)^3|q|^2, \\
\label{sol-A}
f_i(q,\epsilon)&=&-(2\pi)^3\frac{\Pi^T_i}{A^T_i},
\end{eqnarray}
where $f_i(q,\epsilon)$ is given by
\begin{equation}
f_i(q,\epsilon)\delta^{ij}=\sqrt{\gamma}g^{ij}\mathcal F_T(q,\epsilon).
\end{equation}
$\Pi^i_T$ is conjugate momentum of $A^T_i$, which is given by
$\Pi^i_T=\partial_r A^T_i$.  To get some of above expressions, we have
used the explicit form of the background metric(\ref{AdS4-metric}).
In the solution(\ref{sol-A}), the index `$i$' is not summed
over. Since $A^T_i$ is transverse, if we suppose the three momentum,
$q_i$ is along $x$ direction, then $A^T_i$ will have two independent
components $A^T_y$ and $A^T_z$.  In this case, the index $i$ in the
solution (\ref{sol-A}) is either $y$ or $z$ and arbitrary linear
combination of these two solutions is not a solution since
equation(\ref{equation-RG-A}) is non-linear.

Finally, we obtain double trace part of the transverse gauge fields in
the effective action $S_B$.  To do that let us first write down the general solution of 
transverse gauge field from the bulk equation of motion(\ref{A-bulk-eom})
\begin{equation}
A^{T}_{i,q}(r)=\mathcal A^{T(0)}_{i,q}\cosh(|q|r)
+\mathcal A^{T(1)}_{i,q}\sinh(|q|r),
\end{equation}
where $\mathcal A^{T(0)}_{i,q}$ and $\mathcal A^{T(1)}_{i,q}$ are
arbitrary $q_i$ dependent vector functions.  Substituting this general
solution in (\ref{equation-RG-A}), we arrive at the radial
flow of double trace part of transverse gauge field
\begin{equation}
\label{U1-conclusive-action}
S_B=\frac{1}{2}\int d^3 q\left( \frac{\sinh(|q|r)+b_{q}\cosh(|q|r)}{\cosh(|q|r)+b_{q}\sinh(|q|r)}\right) A^T_{i,q}A^T_{i,-q},
\end{equation}
where $b_q$ is a momentum dependent constant and we have only stated
the double trace part of $S_B$.

\subsubsection{The Fokker-Planck action and the Langevin dynamics of $U(1)$ gauge fields}
In this section, we carry out stochastic quantization of $U(1)$ vector
fields. Since the classical effective action (\ref{gamma-aaa}) is
comprised of transverse parts of gauge fields only, we suppress
super(sub)script `$T$' from now on, and assume all the fields in this
section are transverse.  Moreover, for the boundary fields,
superscript $(0)$ is used in the previous section, we will suppress
this too and $A_{i,q}$ is just vector fields appearing in stochastic
quantization.
\paragraph{The Fokker-Planck action}
Using the definition of Fokker-Planck action
(\ref{Fokker-Planck-Lagrangian}) and prescription for the classical
action 
\begin{equation}
\label{A-classical-action}
S_c=2\Gamma[A],
\end{equation}
we get
\begin{eqnarray}
\label{A-fp}
S_{FP}=\int^t_{t_0}dt\int d^3 q \left[ \frac{1}{2}(\partial_t A_{i,q})(\partial_t A_{i,-q})+\frac{1}{2}|q|^2 A_{i,q} A_{i,-q} + \frac{1}{4}|q|\delta(0) \right],
\end{eqnarray}
where we have used
\begin{eqnarray}
\frac{\delta S_{cl}[A]} {\delta A_{i,q}}=-2|q|A_{i,-q} {\rm\  \ and \ \ }\frac{\delta^2 S_{cl}[A]}{\delta A^{i}_{q} \delta A_{i,p}}= -2|q|\delta^3(q+p).
\end{eqnarray}
The last term is an infinite constant, and does not contribute to the bulk dynamics. The second term can be manipulated as
\begin{equation}
|q|^2 A_{i,q} A_{i,-q}=-iq_j A_{i,q}  iq_j A_{i,q} -iq_i A_{i,q}  iq_j A_{j,q} = \frac{1}{2} F_{ij,q}F^{ij}_{-q},
\end{equation}
where we have used the fact that the gauge field that appears in the
first equality is transverse. With this the Fokker-Planck action becomes
\begin{equation}
S_{FP}=\frac{1}{4}\int^t_{t_0} dt\int d^3 q F_{\mu\nu,q}F^{\mu\nu}_{-q}.
\end{equation}
This Fokker-Planck Lagrangian density has the same form as bulk Lagrangian density from which the boundary action is obtained. 

To study stochastic time evolution of the action (\ref{A-fp}), let us derive
equations of motion from it.  The equation of motion is given by
\begin{equation}
\label{EOM-ads2-scalarA}
0=\ddot A_{i,q}-q^2 A_{i,q}.
\end{equation}
The most general solution of the equation of motion is
\begin{equation}
A_{i,q}(t)=\bar \mathcal  A_{i,q}\cosh(|q|t)+\tilde\mathcal  A_{i,q}\sinh(|q|t),
\end{equation}
where $\bar\mathcal A_{i,q}$ and $ \tilde\mathcal A_{i,q}$ arbitrary
vector functions of 3-momenta, $q_i$.

We impose the boundary condition by assuming that at certain time $t$,
the gauge field satisfies $A_{i,q}(\tilde t=t)=A_{i,q}(t)$. Then, the solution
becomes
\begin{equation}
\label{final-solution-phi-preA}
A_{i,q}(\tilde t)=A_{i,q}(t)\frac{\cosh(|\omega|\tilde t)+\mathcal B_{i,q}\sinh(|\omega|\tilde t)}{\cosh(|\omega|t)+\mathcal B_{i,q}\sinh(|\omega|t)},
\end{equation}
where $\mathcal B_{i,q}=\frac{\tilde \mathcal A_{i,q}}{\bar \mathcal A_{i,q}}$ and index $i$ is not summed.

{\it At this point, we stress that the Fokker-Planck action contains
  more degrees of freedom. Since, $A_{i,q}(t)$ has two independent
  degrees of freedom $A_{y,q}(t)$, $A_{z,q}(t)$(assuming the momentum
  $q_i$ is along $x$-direction), initial conditions for $A_{y,q}(t)$
  and $A_{z,q}(t)$ will be different, which are determined by choices
  of $\mathcal B_{y,q}$ and $\mathcal B_{z,q}$. However, as argued in
  the last section, holographic renormalization group computation
  contains only a single constant $b_q$ in
  (\ref{U1-conclusive-action}).}\footnote{This distinction occurs
  because the double trace coupling in the holographic renormalization
  group satisfies a non-linear equation (\ref{equation-RG-A}) but
  there is no such obvious condition appearing in the Fokker-Planck,
  and as we will see later, in the Langevin dynamics as well.  It
  would be useful to understand this issue better to develop a closer
  analogy between these two formalisms.} To reproduce this correctly,
we set
\begin{equation}
\label{A-ini-b-con}
d_q\equiv\mathcal B_{y,q}=\mathcal B_{z,q}.
\end{equation}


After this, we plug the solution (\ref{final-solution-phi-preA}) into Fokker-Planck action and evaluate it. It is given by
\begin{eqnarray}
\label{mid-final-form-FPA}
S_{FP}&=&\int^{t}_{-\frac{1}{|q|}coth^{-1}(d_q)}d\tilde t \int d^3q \left(\frac{1}{2}(\partial_t A_{i,q})(\partial_t A_{i,-q})
+\frac{1}{2}|q|^2 A_{i,q} A_{i,-q}  \right) \\ \nonumber
&=&\left.\frac{1}{2}\int d^3 q A_{i,q}(\tilde t)
 \dot A_{i,-q}(\tilde t) \right|^{\tilde t=t}_{\tilde t=-\frac{1}{|q|}\coth^{-1}(d_q)} \\ \nonumber
&=&\frac{1}{2}\int d^3q |q|A_{i,q}(t)  A_{i,-q}(t)
\left( \frac{\sinh(|q| t)+d_q\cosh(|q| t)}{\cosh(|q|t)+d_q \sinh(|q|t)}\right),
\end{eqnarray}
where for the second equality, we have used equation of motion(\ref{EOM-ads2-scalarA}) and the lower limit of the integration has chosen by the same way that we 
have done in Sec.\ref{Stochastic quantization of the classical effective action: Fokker-Planck approach}. 
The Fokker-Planck action(\ref{mid-final-form-FPA}) is the same form with (\ref{U1-conclusive-action}) under the condition that $d_q=b_q$.

\paragraph{The Langevin dynamics} 
We start with Langevin equation from the classical action (\ref{A-classical-action}) 
\begin{equation}
\label{Langevin equation}
\frac{\partial A_{i,q}(t)}{\partial t}= - \frac{\delta S_{cl}[A_{i,q}(t)]}{\delta A_{i,-q}(t)}+ \eta_{i,q}(t)=-|q|A_{i,q}+ \eta_{i,q}(t).
\end{equation}
The solution of this equation with the initial condition prescribed in
Sec.\ref{Stochastic quantization of the classical effective action:
  Fokker-Planck approach} is
\begin{equation}
A^{a}_{i,q}(t)=\int^{t}_{-\frac{1}{|q|}\coth^{-1}(\mathcal B_{i,q})}d \tilde t e^{-|q|(t-\tilde t)}\delta_{ij}\eta_{j,q}(\tilde t)
\end{equation}
where the index $j$ is summed over but index $i$ is free.

{\it The same feature of stochastic quantization arises here too. For each component of gauge fields, one can assign different boundary conditions by choosing 
$\mathcal B_{i,q}$ differently. However, to reproduce holographic renormalization group calculations, we impose the same condition as (\ref{A-ini-b-con}).}

With such a choice, we compute stochastic correlation function using 
\begin{equation}
<\eta_{i,q}(t)\eta_{j,q^\prime}(t^\prime)>=\delta_{ij}\delta^3(q-q^\prime)\delta(t-t^\prime),
\end{equation}
which is given by
\begin{equation}
\label{A-2-point-st}
<A_{i,q}(t)A_{j,q^\prime}(t)>=\delta_{ij}\delta^3(q-q^\prime)\frac{1}{2|q|}\left( 1-\frac{d_q-1}{d_q+1}e^{-2|q|t}  \right).
\end{equation}
(\ref{A-2-point-st}) is consistent with (\ref{U1-conclusive-action}) through the relation(\ref{Langevin-finally}), provided that $d_q=b_q$ and $t=r$.
\section{Conclusion and Open Questions}
\setcounter{equation}{0}
In this paper, we have shown that in the case that the bulk action is Weyl invariant and allows alternative quantization,
stochastic quantization of the classical action which is given by $S_c=2\Gamma$, where $\Gamma$ is classical effective action from the bulk gravity theory
without any deformations precisely captures the radial flow of double trace deformation coupling in holographic Wilsonian renormalization group computation.
We have studied this proposal by analyzing a couple of examples, (minimally coupled)massless scalar field in $AdS_2$ and $U(1)$ gauge fields in $AdS_4$ as bulk theories. In these examples, 
the radial flow of the double trace couplings is precisely obtained from the stochastic time evolution of the corresponding Fokker-Planck action and Langevin dynamics. 

Even if these examples are quite successful, there are many open
questions, some of which we will list here.
\begin{itemize}
 \item In our example, we only dealt with Weyl invariant action. If
   the bulk action is not Weyl invariant, there must be divergent
   pieces in the
near boundary expansion of the bulk solutions. In such cases, one
needs to add counter-terms to cancel divergent contributions to the
boundary on-shell action.  These counter-terms could modify the
identification, $S_c=2\Gamma$. 

\item 
  Not many examples of interacting boundary conformal field theories
  have been studied either in the holographic Wilsonian
  renormalization group method or in the stochastic quantization
  method.  The Langevin dynamics, however, does provide a method to
  deal with interactions in perturbative expansion in small
  coupling\cite{Huffel1}. Nevertheless, application of this method to study
  the relation between holographic Wilsonian RG and stochastic
  quantization is still an open question.  In \cite{Jatkar:2012mm},
  the authors developed boundary theories of $SU(2)$ Yang-Mills in
  $AdS_4$ and which provide boundary effective action with exotic
  momentum dependent interaction vertices. This is a natural extension
  of $U(1)$ theory in $AdS_4$ to add interactions in it and at the
  same time retains some of the merits of the $U(1)$ case: the bulk
  action is still Weyl invariant and allows alternative quantization.
  One might think about stochastic quantization of this boundary
  theory to extend our argument further.

\item The last question is how the relation will be modified if the
  bulk geometry is not pure $AdS$ space. For example, in \cite{Hong1},
  the authors study holographic Wilsonian RG in extremal black brane
  background. In this case, there are emergent $IR$-$CFT$ near black
  brane horizon since the near horizon geometry is $AdS_2$ and there
  will be more than one (non-trivial) $IR$ fixed point.  The question is whether
  stochastic quantization can capture these fixed points
  appropriately.

\item For non-extremal black brane case, we have to deal with
  conformal field theories at finite temperature.  Stochastic noise
  does not provide a notion of `temperature' in the sense that it does
  not correspond to black brane temperature. Even if there is
  stochastic noise, in our examples the corresponding bulk geometry is
  still pure $AdS$.  It will therefore be interesting to figure out
  how a finite temperature system from $AdS/CFT$ would be accommodated
  in our prescription.  For some realted literature, see
  \cite{Furuuchi:2006st}.  A better understanding of this will put
  this proposal on a firmer footing.

\end{itemize}

\subsection*{Acknowledgments} {We would like to thank Ashoke Sen,
  Rajesh Gopakumar, Bom Soo Kim, Taeyoon Moon, Jaehoon Jeong for
  useful discussion.  D.P.J. would like to thank Theory Division, CERN
  for hospitality.  J.-H.Oh would like to thank everyone in
  CQUeST(Sogang University) for hospitality, especially to Bum-Hoon
  Lee for invitation, he also thanks his ${\mathcal W.J.}$
}

\end{document}